\documentclass[12pt,a4paper]{article}

\usepackage{amsmath}
\usepackage{amssymb}
\usepackage{amsfonts}
\usepackage{bbm}
\usepackage{dirtytalk}
\usepackage{braket}
\usepackage{qcircuit}
\numberwithin{equation}{section}

\usepackage{graphicx}
\usepackage{subfig}
\usepackage{float}
\usepackage{tikz}
\usetikzlibrary{intersections}
\usetikzlibrary{patterns}

\usepackage[numbers,sort&compress]{natbib}

\newcommand{\R}{{\mathbb{R}}}
\newcommand{\C}{{\mathbb{C}}}

\newcommand{\pqty}[1]{\left( #1 \right)}
\newcommand{\bqty}[1]{\left[ #1 \right]}
\newcommand{\abs}[1]{\left\lvert #1\right\rvert}

\newcommand{\expval}[1]{\langle #1 \rangle}

\newcommand {\dv}[3][ ]{
  \ifx #1 { }
    \frac{d #2}{d #3}
  \else
    \frac{d^{#1} #2}{d #3^{#1}}
  \fi
}

\newcommand{\p}{\partial}

\title{Self-adjoint Momentum Operator for a Particle Confined in a Multi-Dimensional Cavity}

\author{A.\ Mariani, U.-J.\ Wiese}

\date{\small{Albert Einstein Center for Fundamental Physics, Institute for Theoretical Physics,
University of Bern, Sidlerstrasse 5, CH-3012 Bern, Switzerland}\\[2ex]%
    \normalsize{\today}
}

\begin{document} 

\maketitle

\vspace{-1cm}

\begin{abstract} \normalsize
Based on the recent construction of a self-adjoint momentum operator for a particle confined in a one-dimensional interval, we extend the construction to arbitrarily shaped regions in any number of dimensions. Different components of the momentum vector do not commute with each other unless very special conditions are met. As such, momentum measurements should be considered one direction at a time. We also extend other results, such as the Ehrenfest theorem and the interpretation of the Heisenberg uncertainty relation to higher dimensions. \end{abstract}

\newpage
 
\section{Introduction}

Momentum is a central concept in both classical and quantum physics, and is fundamentally related to translational invariance. It finds numerous applications in both finite and infinite systems where translational invariance is a symmetry. In finite regions of space, translational invariance is most commonly preserved by choosing appropriate boundary conditions such as periodic boundary conditions. However, other choices of boundary conditions which may arise in physical systems of interest (such as for example Dirichlet or Neumann boundary conditions) explicitly break translational invariance. In these systems, one might therefore expect momentum not to be a useful physical quantity. For a classical particle, in the bulk, away from the boundaries, one can ignore the effect of the boundary and thus still employ the usual notion of momentum, but this is more complicated in quantum mechanics. In this case, the boundary introduces strong ultraviolet effects which need to be properly understood. 

Consider a quantum mechanical particle that is confined to a finite region $\Omega \in \R^d$. Then the operator $-i \vec \nabla$ (in units where $\hbar = 1$), which describes the momentum of a particle in the Hilbert space $L^2(\R^d)$ of square-integrable
functions over $\R^d$, is not self-adjoint in $L^2(\Omega)$ (at least if only local physical boundary conditions are imposed). Self-adjointness (and not Hermiticity alone) is essential to ensure that an operator has a spectrum of real eigenvalues and a corresponding system of orthonormal eigenfunctions, at least in a generalized sense \cite{Gelfand}. In addition to Hermiticity (which results if an operator $A$ and its Hermitean conjugate $A^\dagger$ act in the same way), self-adjointness requires that the corresponding domains coincide, $D(A^\dagger) = D(A)$ \cite{VonNeumann32,Ree75,Gieres_2000}. 
The domain of an operator is usually characterized by square-integrability conditions on derivatives of the wave function as well as by boundary conditions, which are characterized by a family of self-adjoint extension parameters \cite{Bal70,Car90,AlHashimiWiese12}. The consistent interpretation of quantum mechanical measurements of an observable $A$, which return one of its eigenvalues and collapse the wave function onto the corresponding eigenfunction, indeed requires that $A$ is self-adjoint. Since the operator $- i \vec \nabla$ is not self-adjoint in $L^2(\Omega)$, the problem is usually considered in $L^2(\R^d)$. Then the unquantized eigenvalues $\vec k \in \R^d$ form a continuous spectrum. Since the corresponding eigenstates are plane waves $\exp(i \vec k \cdot \vec x)$, which exist everywhere in space with the same probability density, a momentum measurement of this type transfers an infinite amount of energy to the particle and catapults it out of the finite region $\Omega$. 

Recently, a self-adjoint momentum operator for a particle that is strictly confined to a finite 1-d interval $\Omega = [-\tfrac{L}{2},\tfrac{L}{2}]$, even after a momentum measurement, has been constructed \cite{alHashimiWieseAltMomentum, alHashimiWieseHalfLine}. 
In that case, the momentum eigenvalues are quantized. The key to this construction is the doubling of the Hilbert space to $L^2(\Omega) \times \C^2$, which was originally motivated by an ultraviolet lattice regularization, and led to a resolution of this long-standing puzzle also directly in the continuum. While sharp impenetrable boundaries are a mathematical idealization, the physical model of the particle in a box and the new momentum concept may be applied as an effective description to many physical systems which are confined inside a limited region of space, such as ultra-cold atoms confined in an optical box trap \cite{UltraColdAtoms}, electrons in a quantum dot \cite{QuantumWellsWiresDots}, domain wall fermions in a higher-dimensional space \cite{Kap92, Sha93} or the phenomenological MIT bag model \cite{MITBagModel1,MITBagModel2, MITBagModel3} for confined quarks and gluons.

Here we extend the construction of the new self-adjoint momentum to higher dimensions. First of all, we carefully consider the problem of the simultaneous measurement of different components of the momentum vector. We find that different components generally do not commute, unless very special conditions are met. In particular, this implies that in a bounded region of space, momentum measurements should only be considered one direction at a time.  Moreover, we show higher-dimensional generalizations of the Ehrenfest theorem and the Heisenberg-Robertson-Schr\"odinger uncertainty relation \cite{Hei27, Ken27, Wey28, Rob29, Sch30, AlHashimiWieseUncertainty, BachelorStudents} for the new momentum operator. 

\section{Boundary Conditions and the New Momentum Concept in Higher Dimensions}
In this section we discuss the boundary conditions which make the Hamiltonian $H$ and the new momentum operator $\vec p_R$ self-adjoint. This extends the discussion of previous works for the analogous one-dimensional operators \cite{alHashimiWieseAltMomentum, BachelorStudents}, to which we point the interested reader for further details.

\subsection{Self-adjoint Hamiltonian}

Let us consider the Hamiltonian
\begin{equation}
    \label{eq:hamiltonian}
    H = - \tfrac{1}{2 m} \Delta + V(\vec x)
\end{equation}
for $\vec x \in \Omega$. We assume that $V(\vec x)$ is a non-singular potential. Performing two partial integrations, one obtains
\begin{align}
    \expval{ H^\dagger \chi|\Psi} &= \expval{\chi|H \Psi}= \nonumber \\ 
    &=\expval{ H \chi|\Psi} + \frac{1}{2 m} \int_{\p\Omega} d\vec n \cdot
    \left[\vec \nabla \chi(\vec x)^* \Psi(\vec x) - 
    \chi(\vec x)^* \vec \nabla \Psi(\vec x)\right] \ .
    \label{HHermiticity}
\end{align}
Hermiticity of $H$ requires the boundary term to vanish. On the other hand, locality requires that the boundary conditions do not relate the values of the wave function (or its derivatives) at different points in space. We impose local Robin boundary conditions
\begin{equation}
    \gamma(\vec x) \Psi(\vec x) + \vec n(\vec x) \cdot \vec \nabla \Psi(\vec x) = 0 \ , \quad \vec x \in \p\Omega \ .
    \label{eq:robinbc}
\end{equation}
Here $\vec n(\vec x)$ is the unit-vector normal to the boundary, which conventionally points outwards. Dirichlet boundary conditions, $\Psi(\vec x) = 0$, correspond to 
$\gamma(\vec x) \rightarrow \infty$, and Neumann boundary condition,  $\vec n(\vec x) \cdot \vec \nabla\Psi(\vec x) = 0$, correspond to $\gamma(\vec x) = 0$.  Wave functions that obey eq.\eqref{eq:robinbc} and whose Laplacian is square-integrable belong to the domain $D(H)$. Inserting eq.\eqref{eq:robinbc} into the boundary term in eq.\eqref{HHermiticity}, the Hermiticity condition reads
\begin{equation}
    \int_{\p\Omega} d^{d-1}x \left[\vec n(\vec x) \cdot \vec \nabla \chi(\vec x)^* + \gamma(\vec x) \chi(\vec x)^*\right] \Psi(\vec x) = 0 \ .
\end{equation}
Since $\Psi(\vec x)$ can still take arbitrary values, this implies
\begin{equation}
    \gamma(\vec x)^* \chi(\vec x) + \vec n(\vec x) \cdot \vec \nabla \chi(\vec x) = 0 \ .
\end{equation}
This characterizes the domain $D(H^\dagger)$ of $H^\dagger$ (which acts on $\chi$). The domains $D(H^\dagger)$ and $D(H)$ coincide only if $\gamma(\vec x)^* = \gamma(\vec x) \in \R$. This defines a family of self-adjoint extensions of $H$, which are characterized by a self-adjoint extension parameter $\gamma(\vec x)$ for each point $\vec x \in \partial\Omega$ in the boundary of the region. The boundary conditions eq.\eqref{eq:robinbc} ensure that the probability current
\begin{equation}
    \label{eq:probability current}
    \vec j(\vec x) = \frac{1}{2 m i} [\Psi(\vec x)^* \vec \nabla \Psi(\vec x) - \vec \nabla \Psi(\vec x)^* \Psi(\vec x)] \ ,
\end{equation}
does not flow outside the region $\Omega$. This follows because its 
perpendicular component $\vec n(\vec x) \cdot \vec j(\vec x)$ obeys
\begin{align}
    \vec n(\vec x) \cdot \vec j(\vec x)=&\frac{1}{2 m i} [\Psi(\vec x)^* \vec n(\vec x) \cdot \vec \nabla \Psi(\vec x) - \vec n(\vec x) \cdot \vec \nabla \Psi(\vec x)^* \Psi(\vec x)] = \nonumber \\
    &=\frac{1}{2 m i} [- \Psi(\vec x)^* \gamma(\vec x) \Psi(\vec x) + \gamma(\vec x)^* \Psi(\vec x)^* \Psi(\vec x)] = 0 \ . 
\end{align}
Self-adjointness is hence directly responsible for ensuring that the particle remains confined in the finite region.

\subsection{Self-adjoint momentum}

While it is straightforward to construct self-adjoint extensions of the 
Hamiltonian, for the operator $A = - i \vec \nabla$ this is not possible in a physically meaningful way. In fact, performing a partial integration, one obtains
\begin{align}
    \expval{ (- i \vec \nabla)^\dagger \chi|\Psi} &= \expval{\chi|(- i \vec \nabla)\Psi} = \nonumber \\ 
    &=\expval{ (- i \vec \nabla) \chi|\Psi}- i \int_{\p\Omega} d^{d-1}x \, \vec{n}(\vec{x}) \chi(\vec x)^* \Psi(\vec x) \ .
    \label{pHermiticity}
\end{align}
Hermiticity requires that the boundary term vanishes. For physical reasons we again limit ourselves to local boundary conditions, which do not relate the wave function values at physically distinct points. Hermiticity then results for Dirichlet boundary conditions, $\Psi(\vec x) = 0$, $\vec x \in \p\Omega$, which define the domain $D(- i \vec \nabla)$. However, as soon as $\Psi(\vec x)$ is fixed to zero at the boundary, $\chi(\vec x)$ can still take arbitrary values. As a consequence, the domain of $(- i \vec \nabla)^\dagger$
(which acts on $\chi(\vec x)$) remains unrestricted, 
$D((- i \vec \nabla)^\dagger) \supset D(- i \vec \nabla)$. Since the two domains do not coincide, although with Dirichlet boundary conditions $- i \vec \nabla$ is Hermitean, it is not self-adjoint. Consequently, it does not qualify as a physically acceptable momentum operator in the Hilbert space $L^2(\Omega)$.

As pointed out recently \cite{alHashimiWieseAltMomentum, alHashimiWieseHalfLine}, a physically and mathematically satisfactory momentum operator can be defined in a doubled Hilbert space with 2-component wave functions
\begin{equation}
    \vec p_R = - i \left(\begin{array}{cc} 0 & \vec \nabla \\ \vec \nabla & 0 
    \end{array}\right) = - i \sigma_1 \vec \nabla \ , \quad \quad
    \Psi(\vec x) = \left(\begin{array}{c} \Psi_e(\vec x) \\ \Psi_o(\vec x) 
    \end{array}\right).
    \label{eq:2component}
\end{equation}
Besides $\vec p_R$, the momentum operator $\vec p = \vec p_R + i \vec p_I$ also has an anti-Hermitean contribution $i \vec p_I$, which we will discuss later.
First we concentrate on the self-adjointness of $\vec p_R$. Properly speaking $\vec p_R$ is a tuple of operators, and only the momentum in a specific direction is an operator on the Hilbert space. Therefore we consider the momentum in direction $\hat{k}$, that is the operator $\hat{k} \cdot \vec p_R$, which is indeed a map $D(\hat{k} \cdot \vec p_R) \to \mathcal{H}$ from its domain $D(\hat{k} \cdot \vec p_R) \subset \mathcal{H}$ to the Hilbert space $\mathcal{H}$. By partial integration one obtains
\begin{align}
    &\expval{(\hat{k} \cdot \vec p_R)^\dagger \chi|\Psi}  = \expval{\chi|(\hat{k} \cdot \vec p_R) \Psi} 
    = \nonumber \\
    &= \expval{(\hat{k} \cdot \vec p_R) \chi|\Psi} -
    i  \int_{\p\Omega} d^{d-1}x [\chi_e(\vec x)^* \Psi_o(\vec x) + 
    \chi_o(\vec x)^* \Psi_e(\vec x)] (\hat{k} \cdot \vec n(\vec x)) \ .
    \label{eq:pRHermiticity}
\end{align}
If the operator $\hat{k} \cdot \vec p_R$ is to be Hermitean, then the boundary term must vanish. Locality then implies that 
\begin{equation}
    \label{eq:prHermiticity equals zero}
    [\chi_e(\vec x)^* \Psi_o(\vec x) + 
    \chi_o(\vec x)^* \Psi_e(\vec x)] (\hat{k} \cdot \vec n(\vec x))=0 \ .
\end{equation}
As a consequence of eq.\eqref{eq:prHermiticity equals zero}, any boundary condition need only be imposed on the set of points $\vec x$ such that $\hat{k} \cdot \vec n(\vec x) \neq 0$. Thus we have two separate situations. If $\hat{k} \cdot \vec n(\vec x) = 0$ only on a set of isolated points, then the condition eq.\eqref{eq:prHermiticity equals zero} is equivalent to 
\begin{equation}
    \chi_e(\vec x)^* \Psi_o(\vec x) + \chi_o(\vec x)^* \Psi_e(\vec x)=0 
\end{equation}
everywhere on the boundary, and as such, we can now impose the boundary conditions
\begin{equation}
    \Psi_o(\vec x) = \lambda(\vec x) \Psi_e(\vec x) \ , \quad \vec x \in \p\Omega \ ,
    \label{eq:momentumbc}
\end{equation}
which constrain the domain $D(\hat{k}\cdot\vec p_R)$. Inserting these relations in the square bracket in eq.\eqref{eq:prHermiticity equals zero}, the Hermiticity condition takes the form
\begin{equation}
    [\chi_e(\vec x)^* \lambda(\vec x) +  \chi_o(\vec x)^*] \Psi_e(\vec x) = 0 \ , \quad\quad \vec x \in \p\Omega \ .
\end{equation}
Since $\Psi_e(\vec x)$ can take arbitrary values, this implies
\begin{equation}
    \chi_o(\vec x) = - \lambda(\vec x)^* \chi_e(\vec x) \ , \quad \quad \vec x \in \p\Omega  \ .
\end{equation} 
The self-adjointness of $\hat{k}\cdot\vec p_R$ requires $D(\hat{k}\cdot\vec p_R^{\,\dagger}) = D(\hat{k}\cdot\vec p_R)$, which implies $\lambda(\vec x) = - \lambda(\vec x)^*$ such that $\lambda(\vec x) \in i \R$. Hence, the self-adjoint extensions of $\hat{k}\cdot\vec p_R$ are characterized by a purely imaginary parameter $\lambda(\vec x)$ at each point on the boundary. Note however that the parameter $\lambda$ also depends on $\hat{k}$, that is momentum operators in different directions may have different self-adjoint extension parameters.

On the other hand, depending on the choice of domain $\Omega$ and direction $\hat{k}$, it is also possible that $\hat{k} \cdot \vec n(\vec x) = 0$ on a subset of $\partial\Omega$ of non-zero measure. This occurs whenever $\partial\Omega$ includes hyperplanes of codimension $1$, for example straight lines in $2D$, planes in $3D$, etc. When this occurs, for $\hat{k}\cdot\vec p_R$ to be self-adjoint, both $\Psi$ and $\chi$ must satisfy the usual boundary conditions $\Psi_o(\vec x) = \lambda(\vec x) \Psi_e(\vec x)$ with $\lambda(\vec x) \in i \R$ on every point $\vec x \in \partial \Omega$ such that $\hat{k} \cdot \vec n(\vec x) \neq 0$. On the other hand, no boundary conditions at all are imposed at those points $\vec x$ where $\hat{k} \cdot \vec n(\vec x) = 0$.

As explained in previous work \cite{alHashimiWieseAltMomentum, BachelorStudents} in order to recover the original physics from the doubled Hilbert space $L^2(\Omega) \times \C^2$, the Hamiltonian needs to be appropriately modified so that only those states $\ket{\Psi}$ with $\Psi_e=\Psi_o$ should be considered as belonging to the physical, finite-energy subspace.

\section{Momentum measurements in higher dimensions}

In this section we discuss the measurement of the new momentum operator $\vec p_R$ in higher dimensions. The discussion is based on the one-dimensional case which was already considered in previous works \cite{alHashimiWieseAltMomentum, alHashimiWieseHalfLine, BachelorStudents}. We find that different components of the momentum are not simultaneously measurable, unless the region $\Omega$ is a perfect parallelepiped and the self-adjoint extension parameters $\lambda$ are constant on each face. For a general momentum measurement in higher dimensions, one therefore chooses a specific direction. The position operators in all transverse directions are then measured, and, as we will see in this section, the resulting momentum measurement reduces to the one-dimensional case.

Before we start the discussion, we point out a subtlety of the theory of the simultaneous diagonalization of unbounded self-adjoint operators. In fact, in order for two self-adjoint operators $A$ and $B$ to be simultaneously diagonalizable, it turns out not to be enough that $[A,B]=0$ on a dense subset of the Hilbert space where the relation is well-defined \cite{Ree75, Gieres_2000, Juric__2022}. For $A$ and $B$ to be simultaneously diagonalizable they must \textit{strongly commute}, a condition which may be equivalently stated as either the commutation of all (bounded) projection operators occurring in their spectral decomposition \cite{Gieres_2000}, or the commutation of the respective families of one-parameter exponentials for all values of the parameters \cite{Juric__2022}.

\subsection{General structure of momentum measurements}

First of all we note that it is possible to simultaneously measure the position operator and the momentum operator in orthogonal directions. In fact one has
\begin{equation}
    [\hat{k} \cdot \vec p_R, \hat{m} \cdot \vec x] = 0 \quad \quad \mathrm{if} \quad \quad \hat{k} \cdot \hat{m} = 0,
\end{equation}
and this relation is well-defined on the whole domain of $\hat{k} \cdot \vec p_R$. This is because $\hat{m} \cdot \vec x$ is a bounded operator defined on the whole Hilbert space and, moreover, if $\Psi$ satisfies the boundary conditions eq.\eqref{eq:momentumbc}, so does $\hat{m} \cdot \vec x\, \Psi$. 

\begin{figure}
    \centering
    \begin{tikzpicture}
        \draw[name path=blob]  plot[smooth, tension=.7] coordinates {(-3.0,-0.5) (-1,-1.5) (3.0,0) (3.0,2.0) (0,3.5) (-2.0,2.5) (-1.0,1.5) (-3.5,1.0) (-3.0,-0.5)};
        
        \path[name path=line right] (-0.25,-1.5) -- (2.25,3.0);
        \draw[name intersections={of=blob and line right}] (intersection-1) -- (intersection-2);

        \path[name path=line left] (-3.25,-1.0) -- (-0.75,3.5);
        \draw[name intersections={of=blob and line left}] (intersection-2) -- (intersection-3);
        \draw[name intersections={of=blob and line left}] (intersection-1) -- (intersection-4);
    \end{tikzpicture}
    \caption{A non-convex two-dimensional domain $\Omega$ together with its intersection with two parallel lines. The left line is split into two intervals because $\Omega$ is non-convex.}
    \label{fig:momentum measurement 2d}
\end{figure}
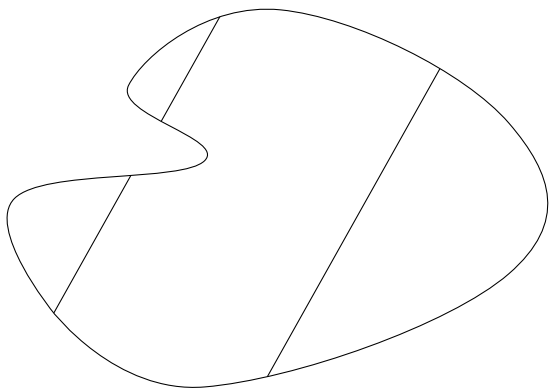

Therefore, the general structure of a higher-dimensional momentum measurement in a certain direction $\hat{k}$ involves first measuring the position operator $\vec x$ in all directions orthogonal to $\hat{k}$. This will single out a line in the $d$-dimensional space, which will pierce through the boundary $\p \Omega$ in a set of isolated points, as shown in Fig.~\ref{fig:momentum measurement 2d} (note that the case when the line is parallel to the boundary does not cause problems, as in that case $\vec{n} \cdot \hat{k} = 0$ and the boundary condition eq.\eqref{eq:momentumbc} does not apply). The simplest case is that of a convex domain $\Omega$, in which the line will generically pierce through $\partial \Omega$ in exactly two points. Then the measurement of the $\hat{k}$ component of the momentum is the same as the measurement of the one-dimensional momentum operator, as described in previous work \cite{alHashimiWieseAltMomentum,alHashimiWieseHalfLine}, with values of $\lambda_{\pm}$ equal to those of $\lambda(\vec x)$ at the two points $\vec{x}_\pm$ pierced by the line. If, on the other hand, the domain is non-convex, then the intersection between the line and $\Omega$ will be a number of disconnected one-dimensional intervals, on each of which the momentum may be measured as in the one-dimensional case. The spectrum will therefore be the union of the single-interval spectra. In both cases, the overall spectrum of $\hat{k} \cdot \vec p_R$ will be the union of the spectra of the measurement $\hat{k} \cdot \vec p_R$ for each possible choice of eigenvalues of the position operator $\vec x$ in all directions orthogonal to $\hat{k}$. Its spectrum will thus be \textit{continuous}, even though it is discrete in each line. We will consider the explicit form of the eigenfunctions in the next section.

\subsection{Simultaneous measurement of the momentum in different directions}

In this section we consider the simultaneous measurement of the momentum operator in orthogonal directions. For simplicity we first consider the two-dimensional case, and rotate the axes so that the two directions coincide with the $\hat{x}$ and $\hat{y}$ axes, i.e. we consider the operators $\hat{x} \cdot \vec p_R$ and $\hat{y} \cdot \vec p_R$. In this case, there are no position operators orthogonal to both the $x$ and $y$ directions. We must consider several cases, in relation with the shape of the box and the choice of self-adjoint extension parameters $\lambda$.

First of all, consider an irregularly shaped domain $\Omega$, which is one where $\vec{n} \cdot \hat{k} = 0$ only at a set of isolated points as $\hat{k}$ ranges among a set of basis unit-vectors. In this case the momentum boundary conditions eq.\eqref{eq:momentumbc} are imposed at every point of the boundary. We call $\lambda_x(\vec x)$ and $\lambda_y(\vec x)$ the self-adjoint extension parameters appearing in eq.\eqref{eq:momentumbc} for $\hat{x} \cdot \vec p_R$ and $\hat{y} \cdot \vec p_R$ respectively, which may be different in principle. Then if $\lambda_x(\vec x) \neq \lambda_y(\vec x)$, the boundary conditions for the two operators are incompatible and, as such, the commutator $[\hat{x}\cdot\vec p_R, \hat{y}\cdot\vec p_R]$ is only defined on the zero vector. Therefore the two operators cannot be simultaneously diagonalized. 

If instead $\lambda_x(\vec x)=\lambda_y(\vec x)$, still in an irregularly shaped $\Omega$, then $\hat{x} \cdot \vec p_R$ and $\hat{y} \cdot \vec p_R$ are defined in the same domain, and therefore there might be hope that they could have a joint set of eigenfunctions. However, their commutator $[\hat{x}\cdot\vec p_R, \hat{y}\cdot\vec p_R]$ is again only defined on the zero vector, because even if $\Psi$ satisfies the boundary conditions eq.\eqref{eq:momentumbc}, generally $\hat{k}\cdot\vec p_R \, \Psi$ will not. In fact eq.\eqref{eq:momentumbc}, being defined only on the boundary of $\Omega$, can only imply conditions on derivatives of $\Psi$ along directions tangent to $\p \Omega$ at each point. Hence knowledge of the values of $\Psi$ at the boundary cannot restrict $\hat{k}\cdot\vec p_R \, \Psi$ in this case since we assumed that $\vec n \cdot \hat{k} \neq 0$. Therefore again it is not possible to simultaneously diagonalize the two operators. 

We can understand the situation more explicitly by considering the eigenvalue equation for $\hat{x} \cdot \vec p_R$ in a two-dimensional region $\Omega$. For simplicity we assume that $\Omega$ is convex. We can solve
\begin{equation}
    (\hat{x} \cdot \vec p_R) \, \Phi = \mu \Phi \ ,
\end{equation}
to obtain the generic eigenfunction
\begin{equation}
    \label{eq:pRx eigenfunction}
    \Phi(x,y) = A(y) \begin{pmatrix} e^{i \mu x} + \sigma(y) e^{-i \mu x} \\ e^{i \mu x} - \sigma(y) e^{-i \mu x} \end{pmatrix} \ ,
\end{equation}
which should be compared to the one-dimensional case \cite{alHashimiWieseAltMomentum}. Imposing the boundary conditions eq.\eqref{eq:momentumbc}, we see that we must have 
\begin{equation}
    e^{2i \mu x} = \sigma(y) \frac{1+\lambda(x,y)}{1-\lambda(x,y)} \ ,
\end{equation}
at each point $(x,y) \in \p \Omega$. Since we assumed that the surface is convex, each line of fixed $y$ intersects $\p \Omega$ in exactly two points, which we call $(x_-,y)$ and $(x_+,y)$. Then the solution of the eigenvalue equation is similar to the one-dimensional case, whereby
\begin{equation}
    \label{eq:pRx eigenvalue condition}
    e^{2i\mu(x_+-x_-)} = \frac{(1+\lambda(x_+,y))(1-\lambda(x_-,y))}{(1-\lambda(x_+,y))(1+\lambda(x_-,y))}  \ ,
\end{equation}
which gives the spectrum of eigenvalues at each fixed $y$. Since this would imply that the eigenvalue $\mu$ is a function of $y$, we must then choose $A(y) \propto \delta(y-y_0)$ where $\delta$ is the Dirac delta function and $y_0$ a generic value of $y$. Therefore each eigenfunction of $(\hat{x} \cdot \vec p_R)$ is labelled by an eigenvalue $y_0$ of $y$ and then by an integer $n$, which indexes the discrete spectrum at each $y_0$. These results agree with the general measurement prescription that was described at the beginning of the section.

It is clear that the function eq.\eqref{eq:pRx eigenfunction} cannot be an eigenfunction of $\hat{y} \cdot \vec p_R$, the self-adjoint momentum operator in the $y$ direction. In fact, as we see from condition \eqref{eq:pRx eigenvalue condition}, for eq.\eqref{eq:pRx eigenfunction} to be an eigenfunction of $\hat{y} \cdot \vec p_R$, it is necessary not only that $\lambda$ be constant on the whole $\p \Omega$, but also that the distance $x_+-x_-$ be independent of $y$. Since this requirement applies not only for the distances $x_+-x_-$ with respect to $y$, but also for the distances $y_+-y_-$, which would arise in an eigenvalue equation for $\hat{y} \cdot \vec p_R$, this means that the region $\Omega$ in this case must be the inside of a parallelogram in two dimensions, or a parallelepiped in $d$-dimensions. This, however, means that $\vec n \cdot \hat{k} = 0$ for some $\hat{k}$ on the whole domain, a situation which we treat in the next paragraph.

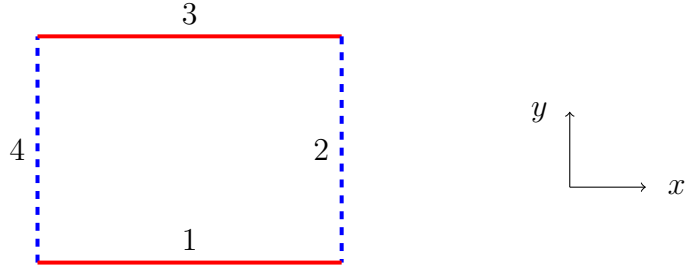
\begin{figure}
    \centering
        \begin{tikzpicture}
        \begin{scope}
            \newcommand\x{1.5pt}
            \draw[dashed, blue, line width=\x] (0,0) -- (0,3) node[midway, left, black] {$4$};
            \draw[red, line width=\x] (0,3) -- (4,3) node[midway, above, black] {$3$};
            \draw[dashed, blue, line width=\x] (4,3) -- (4,0) node[midway, left, black] {$2$};
            \draw[red, line width=\x] (4,0) -- (0,0) node[midway, above, black] {$1$};
        \end{scope}
        \begin{scope}[xshift=5cm]
            \draw[->] (2,1) -- (3,1) node[label=right:{$x$}] {};
            \draw[->] (2,1) -- (2,2) node[label=left:{$y$}] {};
        \end{scope}
        \end{tikzpicture}

    \caption{A two-dimensional rectangle, an example of a parallelepiped, where the boundary conditions of the self-adjoint momentum in orthogonal directions are applied on disjoint subsets. The segments where $\hat{x} \cdot \vec p_R$ satisfies the boundary conditions eq.\eqref{eq:momentumbc} are dashed (blue), while the segments where $\hat{y} \cdot \vec p_R$ satisfies the same boundary conditions are solid (red).}
    \label{fig:rectangle 2D}
\end{figure}

The final case to consider is when $\vec{n} \cdot \hat{k}$ is allowed to vanish on part of $\p \Omega$. Again we consider for simplicity a two-dimensional region $\Omega$ with the operators $(\hat{x} \cdot \vec p_R)$ and $(\hat{y} \cdot \vec p_R)$. We recall that on those points where $\vec{n} \cdot \hat{k} = 0$ no boundary conditions are imposed on $\hat{k} \cdot \vec p_R$. From the previous discussion, it is clear that as soon as we must impose boundary conditions for both $(\hat{x} \cdot \vec p_R)$ and $(\hat{y} \cdot \vec p_R)$ on a subset of $\p \Omega$ of non-zero measure, then the domain of the commutator $[\hat{x}\cdot\vec p_R, \hat{y}\cdot\vec p_R]$ is only the zero vector, and therefore the two operators cannot be simultaneously diagonalized. In general, this means that a necessary condition for the two operators to commute is that the domain $\p \Omega$ must be the union of pairwise parallel hyperplanes, so that $\Omega$ is the interior of a $d$-dimensional parallelepiped. In two dimensions this means that $\Omega$ must be a parallelogram in general, or, in the case of the operators $(\hat{x} \cdot \vec p_R)$ and $(\hat{y} \cdot \vec p_R)$, a rectangle, as shown in Fig.~\ref{fig:rectangle 2D}. This is the same conclusion that we reached by considering the eigenvalue equation for $(\hat{x} \cdot \vec p_R)$ in the previous paragraph. It should be noted, however, that if the one or more of the corners of the parallelepiped are rounded off, then we would have to impose boundary conditions for multiple components of $\vec p_R$ simultaneously. Their commutator would then again be defined only on the zero vector, so that even an infinitesimal rounding off of the corners of the square makes it impossible to simultaneously measure the two operators. 

We now investigate in more detail the case of the operators $(\hat{x} \cdot \vec p_R)$ and $(\hat{y} \cdot \vec p_R)$ on a two-dimensional rectangle, shown in Fig.~\ref{fig:rectangle 2D}. In this case the momentum in the $x$ direction, $\hat{x} \cdot \vec p_R$, only has boundary conditions on the line segments parallel to the $y$ axis (for which $\vec n \propto \hat{x}$, so that $\vec{n} \cdot \hat{x} \neq 0$), while $\hat{y} \cdot \vec p_R$ only has boundary conditions on the line segments parallel to the $x$ axis. As we've already seen, $\lambda$ must be constant on each line segment of $\p \Omega$. The domain of the commutator $[\hat{x}\cdot\vec p_R, \hat{y}\cdot\vec p_R]$ is now non-zero, and given by those square-integrable wave functions $\Psi$ in the doubled Hilbert space with square-integrable derivatives such that
\begin{equation}
    \begin{cases}
        \begin{matrix}
            \Psi_o(\vec{x}) = \lambda_{1,3} \Psi_e(\vec{x}) \ ,\\
            \p_y\Psi_o(\vec{x}) = \lambda_{1,3} \p_y \Psi_e(\vec{x}) \ ,
        \end{matrix} & \mathrm{on}\,\,\p\Omega_{1,3} \ ,\\
        \begin{matrix}
            \Psi_o(\vec{x}) = \lambda_{2,4} \Psi_e(\vec{x}) \ , \\
            \p_x\Psi_o(\vec{x}) = \lambda_{2,4} \p_x \Psi_e(\vec{x}) \ ,
        \end{matrix} & \mathrm{on}\,\,\p\Omega_{2,4} \ ,
    \end{cases}
\end{equation}
where the labels $1$ to $4$ for the line segments making up the boundary $\p\Omega$ are given in Fig.~\ref{fig:rectangle 2D}. Note that we're allowed to differentiate the boundary condition eq.\eqref{eq:momentumbc} in a tangent direction. The domain of $[\hat{x}\cdot\vec p_R, \hat{y}\cdot\vec p_R]$ is therefore a dense subset of the Hilbert space, and moreover $[\hat{x}\cdot\vec p_R, \hat{y}\cdot\vec p_R]=0$ on its whole domain of definition. However, as we have seen in the introduction to this section, this does not conclusively show that the two operators are simultaneously diagonalizable. However, we can show that this is the case by considering the simultaneous eigenvalue equations for both components of $\vec p_R$,
\begin{equation}
    \begin{cases}
        (\hat{x} \cdot \vec p_R) \, \Phi = \mu_x \Phi \ ,\\
        (\hat{y} \cdot \vec p_R) \, \Phi = \mu_y \Phi \ .
    \end{cases}
\end{equation}
The general solutions of the two equations is given by
\begin{equation}
    \label{eq:simultaneous eigenfunction}
    \Phi(x,y) = \begin{pmatrix} A e^{i \vec{\mu}\cdot \vec x} + B e^{-i \vec{\mu}\cdot \vec x} \\ A e^{i \vec{\mu}\cdot \vec x} - B e^{-i \vec{\mu}\cdot \vec x} \end{pmatrix} \ ,
\end{equation}
where $A, B, \vec\mu=(\mu_x, \mu_y)$ are constants. The lengths of the sides of the rectangle are $L_3=L_1=L_x$ and $L_4=L_2=L_y$. The boundary conditions require that
\begin{align}
    \Psi_o(x,0) &= \lambda_1 \Psi_e(x,0) \ , \\
    \Psi_o(L_x,y) &= \lambda_2 \Psi_e(L_x,y) \ , \\
    \Psi_o(x,L_y) &= \lambda_3 \Psi_e(x,L_y) \ , \\
    \Psi_o(0,y) &= \lambda_4 \Psi_e(0,y) \ .
\end{align}
In this case, all the position-dependent factors drop out and this system is therefore identical to two copies of one-dimensional self-adjoint momentum operators, one with overall length $L_x$ and extension parameters $(\lambda_2, \lambda_4)$, the other one with length $L_y$ and extension parameters $(\lambda_1, \lambda_3)$. The two operators may thus be simultaneously diagonalized.

At the end of this section, we summarize our results about the simultaneous diagonalization of different components of the self-adjoint momentum. We found that in general this is not possible unless the region $\Omega$ is a parallelepiped, in which case one may simultaneously measure the components of $\vec p_R$ parallel to the boundaries. However this situation is unphysical, as even an infinitesimal rounding off of the corners of the parallelepiped would make it impossible to simultaneously measure both. The eigenstates of $\hat{l} \cdot \vec p_R$ in a convex region are given by
\begin{equation}
    \label{eq:normalized pR eigenfunction}
    \Phi_{\vec{y}_0,k}(x_l,\vec{y}) = \delta(\vec{y}-\vec{y}_0) \frac{1}{2\sqrt{x_{l+}-x_{l-}}} \begin{pmatrix} e^{i x_l k} + \sigma_{\vec{y}_0,k} e^{-i x_l k} \\ e^{i x_l k} - \sigma_{\vec{y}_0,k}  e^{-i x_l k} \end{pmatrix} \ ,
\end{equation}
where $x_{l+}-x_{l-}$ is the length of the line of constant $\vec{y}_0$ with $\Omega$. The eigenvalues $k$ are implicitly dependent on $\vec{y}_0$ and they are given by
\begin{equation}
    \label{eq:pR eigenvalues}
    k_n = \frac{\pi }{x_{l+}-x_{l-}} n + \theta \ ,
\end{equation}
where $\theta$ is the solution of the equation
\begin{equation}
    \exp{(2i\theta)} = \frac{(1+\lambda_+)(1-\lambda_-)}{(1-\lambda_+)(1+\lambda_-)} \ .
\end{equation}
Here $\lambda_\pm$ are the values of the self-adjoint extension parameter $\lambda(\vec x)$ at the two points pierced by the line of constant $\vec{y}_0$ and thus also depend implicitly on $\vec{y}_0$. Since the wavefunction  eq.\eqref{eq:normalized pR eigenfunction} is an eigenfunction of the position operator in the transverse directions, it contains a non-normalizable $\delta$-function. This should be interpreted as a generalized eigenfunction, as usual for eigenfunctions of operators with a continuous spectrum such as the position operator.  

If, on the other hand, the finite region $\Omega$ is non-convex, the eigenvectors will depend on the number of finite intervals identified by the intersection between $\Omega$ and the line of constant $\vec{y}_0$. If there is only one finite interval, then the eigenvectors and eigenvalues are again given by eq.\eqref{eq:normalized pR eigenfunction} and eq.\eqref{eq:pR eigenvalues} respectively. Generally, if the intersection identifies $n$ finite intervals, the eigenvectors of $\hat{l} \cdot \vec p_R$ will consist of the union of all the functions which are equal to eq.\eqref{eq:normalized pR eigenfunction} on exactly one of the intervals and zero on the other $n-1$ intervals. The eigenvalues are thus the union of the eigenvalues eq.\eqref{eq:pR eigenvalues} on each of the $n$ intervals, each with its appropriate $\lambda_\pm$. If the eigenvalues are all non-degenerate, then a momentum measurement will only find the particle on one interval at a time. On the other hand, if the eigenvalues are degenerate (this can happen for example if the $\theta$ parameters are equal in the different intervals, and their lenghts are integer multiples of each other), then it is also possible to find the particle in superposition of momentum eigenstates belonging to different intervals.

It is important to note that our discussion shows that, in the higher-dimensional setting, it is only meaningful to consider the momentum in one specific direction only. Finally, we remark that while for simplicity we considered the case of a two-dimensional convex region, it is easy to see that the conclusions generalize to a generically shaped region in any number of dimensions.

\section{Ehrenfest Theorem and Interpretation of the Heisenberg Uncertainty Relation}

In the case of a finite interval in one dimension, it has been shown \cite{BachelorStudents} that, in the physical sector, the expectation value of the new momentum operator $p_R$ can be related to that of the standard momentum via the relation
\begin{equation}
    \expval{-i\p_x} = \expval{p_R} + i \expval{ p_I} \ .
\end{equation}
This relation leads to the position-momentum Ehrenfest theorem,
\begin{equation}
    \dv{\expval{x}}{t} = \expval{p_R} \ .
\end{equation}
The new momentum operator also satisfies a version of the momentum-force Ehrenfest theorem,
\begin{equation}
    \dv{\expval{p_R}}{t} = -\expval{V'} + \expval{F_B} \ ,
\end{equation}
where $F_B$ is a force localized at the boundary of the finite interval. In this section, we generalize the Ehrenfest theorem for the new momentum to the higher-dimensional case.

Moreover, we consider the Heinseberg uncertainty relation. In particular, since measuring the new momentum leads outside the physical sector, a new momentum measurement necessarily transfers an infinite amount of energy to the particle. As a consequence, the variance $\Delta p_R$ is generally infinite \cite{BachelorStudents}. Therefore the uncertainty relation for $\vec{p}_R$ is not physically meaningful. On the other hand, the standard momentum $-i \vec{\nabla}$ is not an observable and therefore its uncertainty relation is also not physically meaningful \cite{AlHashimiWieseUncertainty}. Still, a generalization of the Heisenberg-Robertson-Schr\"odinger uncertainty relation is also valid for non-Hermitean operator such as $-i \vec{\nabla}$ and it can be used to derive a physically meaningful inequality where each term is physically measurable \cite{BachelorStudents}. In this section, we also provide a generalization of this inequality to the higher-dimensional case. 

\subsection{Relations between expectation values}

We now prove the higher-dimensional generalization of the relation  $\expval{-i\p_x} = \expval{p_R} + i \expval{ p_I}$, which has been proven in the one-dimensional case in  \cite{BachelorStudents}. We will use the relation several times in this section, to prove the Ehrenfest theorem as well as for the uncertainty relation. This relation will also serve as a basis to define $\vec{p_I}$ in the higher-dimensional setting.

Consider for simplicity a convex region $\Omega$, so that each line which intersects $\p\Omega$ does so in exactly two points. We consider the expectation value of the operator $-i \hat{m} \cdot \vec\nabla$ in an arbitrary finite-energy state $\ket{\Psi}$. We split the position vector $\vec{x} = (x_m, \vec y)$ into a component $x_m$ parallel to $\hat{m}$ and a basis of components $\vec y$ orthogonal to $\hat{m}$. In order to perform the computation, we expand in a basis $\ket{\Phi_{\vec{y}_0, k}}$ of eigenstates of $\hat{m} \cdot \vec p_R$.  These are characterised first by a choice $\vec{y}_0$ of eigenvector of all the coordinates of the position operator orthogonal to $\hat{m}$. This then defines a line in $d$ dimensional space which is parallel to $\hat{m}$. Due to the convexity assumption on $\Omega$, the line intersects $\p\Omega$ in exactly two points, which reduces the problem to the one-dimensional case for each fixed $\vec{y}_0$ and as such defines a pair of parameters $\lambda_{\pm}$ and the discrete spectrum of eigenvalues $\{k\}$. Expanding in a basis of such eigenstates,
\begin{equation}
    \expval{-i \hat{m} \cdot \vec\nabla} = \int d^{d-1}\vec{y}_0 \, \sum_k \expval{\Psi | \Phi_{\vec{y}_0, k}} \bra{\Phi_{\vec{y}_0, k}} (-i \hat{m} \cdot \vec\nabla) \ket{\Psi} \ .
\end{equation}
Now, since the eigenstates $\bra{\Phi_{\vec{y}_0, k}}$ are $\delta$ functions in the directions orthogonal to $\hat{m}$, both inner products involving $\bra{\Phi_{\vec{y}_0, k}}$ reduce to one-dimensional integrals along the intersection between $\Omega$ and the line defined by $\vec{y}_0$. Hence each sum over $k$ reduces to a one-dimensional problem which was already treated in \cite{BachelorStudents}. Adapting the one-dimensional result to the present situation, we find that 
\begin{multline}
    \sum_k \expval{\Psi | \Phi_{\vec{y}_0, k}} \bra{\Phi_{\vec{y}_0, k}} (-i \hat{m} \cdot \vec\nabla) \ket{\Psi} = \sum_k k \expval{\Psi | \Phi_{\vec{y}_0, k}} \expval{\Phi_{\vec{y}_0, k} | \Psi}+\\
    - \frac{i}{2} \bqty{\abs{\Psi(x_{m+}, \vec{y}_0)}^2-\abs{\Psi(x_{m-}, \vec{y}_0)}^2} \ ,
\end{multline}
where $x_m$ is the coordinate parallel to $\hat{m}$ and $x_{m\pm}$ are the coordinates of the intersection between the line defined by $\vec{y}_0$ and $\p\Omega$. Therefore $x_{m\pm}$ implicitly depend on $\vec{y}_0$. Integrating the first term over $\vec{y}_0$ simply gives the spectral decomposition for $\hat{m} \cdot \vec p_R$, so that overall
\begin{equation}
    \label{eq:almost relation}
    \expval{-i \hat{m} \cdot \vec\nabla} = \expval{\hat{m} \cdot \vec p_R} -\frac{i}{2} \int d^{d-1}\vec{y}_0 \, \bqty{\abs{\Psi(x_{m+}, \vec{y}_0)}^2-\abs{\Psi(x_{m-}, \vec{y}_0)}^2} \ .
\end{equation}
Both points $(x_{m\pm}, \vec{y}_0)$ belong to $\p\Omega$ by construction, and, as such, the last term in eq.\eqref{eq:almost relation} may be written as a difference of two integrals, each of which is performed over half of $\p\Omega$. In fact, the set of points in $\p\Omega$ which are parallel to $\hat{m}$ form a $(d-2)$-dimensional subset which partitions $\p\Omega$ into two disjoint subsets, $\p\Omega_+$ and $\p\Omega_-$ (as long as $\Omega$ is a convex region), such that $\p\Omega = \p\Omega_+ \cup \p\Omega_-$. Since the integration in eq.\eqref{eq:almost relation} is performed over lines parallel to $\hat m$, when integrated over $\p\Omega$ the integrand will be proportional to $\vec n \cdot \hat{m}$, so that
\begin{equation}
    \label{eq:px pR pI relation}
    \expval{-i \hat{m} \cdot \vec\nabla} = \expval{\hat{m} \cdot \vec p_R} -\frac{i}{2} \int_{\p\Omega} d^{d-1}\vec x\, (\vec{n} \cdot \hat{m}) \abs{\Psi}^2 \ ,
\end{equation}
or, equivalently,
\begin{equation}
    \label{eq:px pR pI relation equivalent}
    \expval{-i \vec\nabla} = \expval{\vec p_R} -\frac{i}{2} \expval{ \vec{n} }_{\p\Omega} \ .
\end{equation}
If we interpret the right-hand side of eq.\eqref{eq:px pR pI relation} as the expectation value of an operator $\hat{m} \cdot \vec p_I$, then we can take
\begin{equation}
    \label{eq:pI definition}
    \vec p_I = \lim_{\epsilon \to 0} \begin{pmatrix} -\vec{n}(\vec x) \delta(\vec{x} \in \p\Omega_\epsilon) & 0 \\ 0 & 0 \end{pmatrix} \ ,
\end{equation}
where $\p\Omega_\epsilon$ is a $(d-1)$-dimensional subset of $\Omega$ such that $\lim_{\epsilon \to 0} \p\Omega_\epsilon = \p\Omega$. Note that the factor of $1/2$ in eq.\eqref{eq:px pR pI relation equivalent} is absent in eq.\eqref{eq:pI definition} as it comes from the normalization of the finite-energy state in the doubled Hilbert space.

\subsection{Ehrenfest theorem}

Using the relation eq.\eqref{eq:px pR pI relation}, we may therefore prove the position-momentum Ehrenfest theorem, similarly to what was done in \cite{BachelorStudents} for the one-dimensional case. We start by computing
\begin{equation}
    \label{eq:position exp derivative}
    \dv{}{t}\expval{\hat{k} \cdot \vec x} = i \pqty{\bra{H\Psi}(\hat{k} \cdot \vec x)\ket{\Psi} - \bra{\Psi}(\hat{k} \cdot \vec x)\ket{H \Psi}} \ .
\end{equation}
Using Green's second identity, it is not hard to show that the right hand side of eq.\eqref{eq:position exp derivative} is given by
\begin{multline}
    -\frac{i}{m} \int_\Omega d^d \vec{x}\, \Psi^* (\hat{k} \cdot \vec \nabla) \Psi - \frac{i}{2m} \int_{\p\Omega} d^{d-1} \vec{x}\, (\hat{k} \cdot \vec x) \vec{n} \cdot \bqty{\Psi \vec \nabla \Psi^*-\Psi^* \vec \nabla \Psi }+\\ +\frac{i}{2m} \int_{\p\Omega} d^{d-1} \vec{x}\,(\vec n \cdot \hat k) \abs{\Psi}^2 \ .
\end{multline}
The second term in the first line vanishes because of the Robin boundary conditions eq.\eqref{eq:robinbc}, so that overall
\begin{equation}
    m \dv{}{t}\expval{\hat{k} \cdot \vec x} = \expval{-i \hat{k} \cdot \vec\nabla} + \frac{i}{2} \int_{\p\Omega} d^{d-1}\vec x\, (\vec{n} \cdot \hat{k}) \abs{\Psi}^2 \ .
\end{equation}
Using eq.\eqref{eq:px pR pI relation} it is therefore easy to see that
\begin{equation}
    m \dv{}{t}\expval{\hat{k} \cdot \vec x} = \expval{\hat{k} \cdot \vec p_R} \ ,
\end{equation}
or, equivalently,
\begin{equation}
    m \dv{}{t}\expval{\vec x} = \expval{\vec p_R} \ ,
\end{equation}
as expected. This result reinforces our arguments that $\vec p_R$ is the appropriate concept of momentum for a particle confined to a finite region. 

We now consider the momentum-force Ehrenfest theorem. In this case the calculation does not easily reduce to the one-dimensional case. Therefore we choose to carefully compute $\dv{}{t}\expval{-i \vec \nabla}$ and then take the real part to extract $\dv{}{t}\expval{\vec p_R}$ using eq.\eqref{eq:px pR pI relation}. We have,
\begin{multline}
    \dv{}{t}\expval{-i\hat{k} \cdot \vec \nabla} = -\frac{i}{2m} \pqty{\bra{\Delta\Psi}(-i\hat{k} \cdot \vec \nabla)\ket{\Psi} - \bra{\Psi}(-i\hat{k} \cdot \vec \nabla)\ket{\Delta \Psi}}+\\
    +i\pqty{\bra{V\Psi}(-i\hat{k} \cdot \vec \nabla)\ket{\Psi} - \bra{\Psi}(-i\hat{k} \cdot \vec \nabla)\ket{V \Psi}} \ ,
\end{multline}
where $\Delta$ is the Laplacian. The term in the second line involving the potential may be computed in a straightforward manner and reduces to $-\expval{\hat{k}\cdot\vec\nabla V}$. The term on the right-hand side of the first line, however, is more complicated. In fact, while we may assume that $\Psi$ is twice differentiable as it is in the domain of the Hamiltonian, it may not be thrice differentiable. Therefore in the inner product $\bra{\Psi}(-i\hat{k} \cdot \vec \nabla)\ket{\Delta \Psi}$, the operator $(-i\hat{k} \cdot \vec \nabla)$ should be understood as acting on the left. We therefore perform a partial integration to show that
\begin{multline}
    \bra{\Delta\Psi}(\hat{k} \cdot \vec \nabla)\ket{\Psi} - \bra{\Psi}(\hat{k} \cdot \vec \nabla)\ket{\Delta \Psi}=\int_{\Omega} d^d \vec{x}\, \bqty{\nabla^2 \Psi^* (\hat{k} \cdot \vec \nabla \Psi)+\nabla^2\Psi (\hat{k} \cdot \vec \nabla \Psi^*)} +\\
    -\int_{\p\Omega} d^{d-1} \vec{x}\, (\vec{n} \cdot \hat{k}) \Psi^* \nabla^2 \Psi \ .
\end{multline}
Now we would like to show that the remaining volume integral is a boundary term. By careful use of Green's identities, making sure that only two derivatives act on $\Psi$ at any point, we find that
\begin{equation}
    \nabla^2 \Psi^* (\hat{k} \cdot \vec \nabla \Psi)+\nabla^2\Psi (\hat{k} \cdot \vec \nabla \Psi^*)=\vec\nabla\cdot\bqty{(\hat{k}\cdot\vec\nabla\Psi^*)\vec\nabla\Psi+(\hat{k}\cdot\vec\nabla\Psi)\vec\nabla\Psi^* -\hat{k} \vec \nabla \Psi^* \cdot \vec \nabla \Psi} \ ,
\end{equation}
which therefore reduces to a boundary term once substituted back in the volume integral. Then, putting everything together and using the Robin boundary conditions eq.\eqref{eq:robinbc}, we finally find
\begin{multline}
    \dv{}{t}\expval{-i\hat{k} \cdot \vec \nabla}=-\expval{\hat{k}\cdot\vec\nabla V}+\\
    +\frac{1}{2m}\int_{\p\Omega} d^{d-1} \vec{x}\,\bqty{\gamma \hat{k} \cdot \vec \nabla (\Psi \Psi^*)+(\vec{n} \cdot \hat{k})(\vec \nabla \Psi^* \cdot \vec \nabla \Psi+ \Psi^* \nabla^2 \Psi)} \ .
\end{multline}
This expression provides a higher-dimensional generalization of a result first shown in \cite{EhrenfestAlonsoVicenzoGonzalezDiaz}. The imaginary part of this expression gives the expectation value of $\vec p_I$,
\begin{equation}
    \dv{}{t}\expval{\hat{k} \cdot \vec p_I}=
    \frac{1}{4mi}\int_{\p\Omega} d^{d-1} \vec{x}\,\bqty{(\vec{n} \cdot \hat{k})(\Psi^* \nabla^2 \Psi-\Psi \nabla^2 \Psi^*)} \ .
\end{equation}
This can be expressed in terms of the divergence of the probability current $\vec j$, 
\begin{equation}
    \dv{}{t}\expval{\hat{k} \cdot \vec p_I}=
    \frac{1}{2}\int_{\p\Omega} d^{d-1} \vec{x}\,(\vec{n} \cdot \hat{k}) \vec\nabla \cdot \vec j = -\frac{1}{2}\int_{\p\Omega} d^{d-1} \vec{x}\,(\vec{n} \cdot \hat{k}) \frac{\p}{\p t} \abs{\Psi}^2 \ ,
\end{equation}
where we used the continuity equation. This last equation can immediately be seen to be true from the definition of $\vec p_I$, eq.\eqref{eq:pI definition}. Finally, taking the real part instead, we find the momentum-force Ehrenfest theorem,
\begin{multline}
    \dv{}{t}\expval{\hat{k} \cdot \vec p_R}=-\expval{\hat{k}\cdot\vec\nabla V}
    +\\+\frac{1}{2m}\int_{\p\Omega} d^{d-1} \vec{x}\,\bqty{\gamma \hat{k} \cdot \vec \nabla (\Psi \Psi^*)+(\vec{n} \cdot \hat{k})(\vec \nabla \Psi^* \cdot \vec \nabla \Psi+ \Psi^* \nabla^2 \Psi + \Psi \nabla^2 \Psi^*)} \ ,
\end{multline}
or, more simply,
\begin{equation}
    \label{eq:momentum force ehrenfest theorem}
    \dv{}{t}\expval{\vec p_R}=-\expval{\vec\nabla V}
    +\frac{1}{2m}\int_{\p\Omega} d^{d-1} \vec{x}\,\bqty{\gamma \vec \nabla (\Psi \Psi^*)+\vec{n} (\nabla^2(\Psi\Psi^*)-\vec \nabla \Psi^* \cdot \vec \nabla \Psi)} \ .
\end{equation}
Therefore, apart from the usual force term coming from the expectation value of the derivative of the potential, we also have two terms associated with a force at the boundary of quantum mechanical origin. One of these is normal to the boundary and may be interpreted as a sort of quantum-mechanical pressure, while the other one is in the direction of the gradient of the probability density at the boundary and it may be interpreted as arising from the imposition of the Robin boundary conditions eq.\eqref{eq:robinbc}.

\subsection{Interpretation of the Heisenberg Uncertainty Relation}

As shown in a previous work \cite{BachelorStudents}, in its most general form the Heisenberg-Robertson-Schr\"odinger uncertainty relation, valid for not necessarily Hermitean operators $A$ and $B$ \cite{BachelorStudents}, is given by
\begin{equation}
    \label{eq:general inequality}
    \Delta A \Delta B \geq \abs{\expval{A^\dagger  B} - \expval{A^\dagger} \expval{B}} \ .
\end{equation}
In \cite{BachelorStudents} eq.\eqref{eq:general inequality} was used to provide an interpretation for the Heisenberg uncertainty relation for the non-self-adjoint operator $-i\p_x$. Here we extend this result to the higher-dimensional case, and provide an interpretation for the uncertainty relation for the higher-dimensional operator $- i \vec \nabla$. Setting $A_k = -i \hat{k} \cdot \vec \nabla$, by partial integration one obtains
\begin{align}
    \langle A_k^\dagger A_k \rangle&=\int_\Omega d^dx (-i \hat{k} \cdot \vec \nabla \Psi(\vec x))^*
     (-i \hat{k} \cdot \vec \nabla \Psi(\vec x)) \nonumber \\
    &=-\int_\Omega d^dx \ \Psi(\vec x)^* (\hat{k} \cdot \vec{\nabla})^2 \Psi(\vec x) +
    \int_{\p\Omega} d^{d-1}x\,(\vec n \cdot \hat{k}) \Psi(\vec x)^* \hat{k}\cdot \vec \nabla \Psi(\vec x) \nonumber \\
    &=\expval{ -(\hat{k} \cdot \vec{\nabla})^2} + \expval{ (\vec n \cdot \hat{k}) (\hat{k}\cdot \vec \nabla) }_{\p\Omega} \ .
\end{align}
Therefore choosing a set of basis vectors $\hat{k}$, we find, using the Robin boundary conditions eq.\eqref{eq:robinbc},
\begin{equation}
    \sum_{\hat{k}} \langle A_k^\dagger A_k \rangle = \expval{-\Delta} - \expval{\gamma}_{\p\Omega} \ ,
\end{equation}
where $\Delta$ is the Laplacian and $\gamma$ the self-adjoint extension parameter for the Hamiltonian. Setting $B_m = \hat{m} \cdot \vec{x}$, we can similarly compute
\begin{align}
    \expval{B_m^\dagger A_k }&=\int_\Omega d^dx \ \Psi(\vec x)^* (\hat{m} \cdot \vec{x}) (- i \hat{k} \cdot \vec \nabla) \Psi(\vec x) \nonumber \\
    &=\int_\Omega d^dx \ (- i \hat{k} \cdot \vec \nabla \Psi(\vec x))^* (\hat{m} \cdot \vec x) \Psi(\vec x) + i (\hat{k} \cdot \hat{m}) \int_\Omega d^dx \ |\Psi(\vec x)|^2 + \nonumber \\
    &\quad\quad\quad\quad\quad\quad - i \int_{\p\Omega} d^{d-1}x \, (\vec n \cdot \hat{k}) (\hat{m} \cdot \vec x) \abs{\Psi(\vec x)}^2 \nonumber \\
    &=\expval{ A_k^\dagger B_m } + i (\hat{k} \cdot \hat{m}) - 
    i \expval{ (\vec n \cdot \hat{k}) (\hat{m} \cdot \vec x)}_{\p\Omega} \ ,
\end{align}
Also,
\begin{align}
    \expval{A_k}&=\int_\Omega d^dx \ \Psi(\vec x)^* (- i \hat{k} \cdot \vec \nabla) \Psi(\vec x) \nonumber \\
    &=\int_\Omega d^dx \ (- i \hat{k} \cdot \vec \nabla \Psi(\vec x))^* \Psi(\vec x) -i \int_{\p\Omega} d^{d-1}x \, (\vec n \cdot \hat{k}) \ \abs{\Psi(\vec x)}^2 \nonumber \\
    &=\expval{ A_k^\dagger } - i \expval{ \vec n \cdot \hat{k} }_{\p\Omega} \ ,
\end{align}
consistently with eq.\eqref{eq:px pR pI relation equivalent}. Moreover,
\begin{equation}
    (\Delta A_k)^2 = \expval{A_k^\dagger A_k} - \expval{A_k^\dagger} \expval{A_k} = 2m \expval{T_k} + \expval{ (\vec n \cdot \hat{k}) (\hat{k}\cdot \vec \nabla) }_{\p\Omega} - \expval{\hat{k} \cdot \vec p_R}^2-\expval{\hat{k} \cdot \vec p_I}^2 \ ,
\end{equation}
where $T_k = - \frac{1}{2m} (\hat{k} \cdot \nabla)^2$ is the kinetic energy in the $\hat{k}$-direction, and we used eq.\eqref{eq:px pR pI relation}. We must finally express $\expval{B_m^\dagger A_k }$ in terms of measurable quantities. To do so, we consider the expectation value $\expval{(\hat{m} \cdot \vec x)(-i \hat{l} \cdot \vec \nabla)}$ and insert a complete basis of eigenstates of $\hat{l} \cdot \vec p_R$ to find,
\begin{equation}
    \expval{(\hat{m} \cdot \vec x)(-i \hat{l} \cdot \vec \nabla)} = \int d^{d-1}\vec{y}_0 \, \sum_{\mu} \expval{\Psi | \Phi_{\vec{y}_0, \mu}} \bra{\Phi_{\vec{y}_0, \mu}} (\hat{m} \cdot \vec x) (-i \hat{l} \cdot \vec \nabla) \ket{\Psi} \ .
\end{equation}
Here $\ket{\Phi_{\vec{y}_0, \mu}}$ is an eigenfunction of $\hat{l} \cdot \vec p_R$ with eigenvalue $\mu$, as defined in eqs.\eqref{eq:normalized pR eigenfunction} and \eqref{eq:pR eigenvalues}. Now we compute the second inner product explicitly,
\begin{equation}
    \bra{\Phi_{\vec{y}_0, \mu}} (\hat{m} \cdot \vec x) (-i \hat{l} \cdot \vec \nabla) \ket{\Psi} = \int_{\Omega} d^d \vec{x} \, \Phi_{\vec{y}_0, \mu,+}^* (\hat{m} \cdot \vec x) (-i \hat{l} \cdot \vec \nabla) \Psi \ ,
\end{equation}
where $\Phi_{\vec{y}_0, \mu,+}$ is the projection of $\Phi_{\vec{y}_0, \mu}$ onto the positive energy subspace. Since $\Phi$ is an eigenstate of $\hat{l} \cdot \vec p_R$, its projection satisfies $(-i \hat{l} \cdot \vec \nabla) \Phi_{\vec{y}_0, \mu,+} = \mu \Phi_{\vec{y}_0, \mu,+}$. 
Since $\Phi_{\vec{y}_0, \mu,+}^*$ is only supported on a line where $\vec x = (x_l, \vec{y}_0)$ and $x_{l-} < x_l < x_{l+}$, where $x_l = \hat{l} \cdot \vec{x}$ and $x_{l\pm}$ are the points where the line intersects $\p\Omega$, we find that actually
\begin{equation}
    \bra{\Phi_{\vec{y}_0, \mu}} (\hat{m} \cdot \vec x) (-i \hat{l} \cdot \vec \nabla) \ket{\Psi} = \int_{x_{l-}}^{x_{l+}} dx_l \, \Phi_{\vec{y}_0, \mu,+}^*(x_l) (\hat{m} \cdot \vec x) (-i \p_l)\Psi(x_l, \vec{y}_0) \ .
\end{equation}
This last equation can be simplified by integrating by parts and then performing some straightforward manipulations. Putting everything together, we find that
\begin{multline}
    \expval{(\hat{m} \cdot \vec x)(-i \hat{l} \cdot \vec \nabla)} = \expval{(\hat{l} \cdot \vec p_R) (\hat{m} \cdot \vec{x})}+i (\hat{l} \cdot \hat{m})+\\
    -i\int d^{d-1}\vec{y}_0 \, (\hat{m} \cdot \vec x) \Psi(x_l, \vec{y}_0) \sum_\mu \expval{\Psi | \Phi_{\vec{y}_0, \mu}} \Phi_{\vec{y}_0, \mu,+}^*(x_l)  \big \lvert_{x_{l-}}^{x_{l+}} \ .
\end{multline}
The sum over the eigenvalues $\mu$ is then the same as in the one-dimensional case and may be evaluated as in that case. We see from eq.\eqref{eq:normalized pR eigenfunction} that
\begin{equation}
    \Phi_{\vec{y}_0,\mu,+}(x_l) = \frac{e^{i x_l \mu}}{\sqrt{x_{l+}-x_{l-}}} \ .
\end{equation}
Therefore,
\begin{equation}
    \sum_\mu \expval{\Psi | \Phi_{\vec{y}_0, \mu}} \Phi_{\vec{y}_0, \mu,+}^*(x_l)  \bigg \lvert_{x_{l-}}^{x_{l+}} = \int_{x_{l-}}^{x_{l+}} d\widetilde{x}_l\, \Psi^*(\widetilde{x}_l, \vec{y}_0) \frac{\sum_{\mu} e^{i (\widetilde{x}_l-x_{l}) \mu}}{2(x_{l+}-x_{l-})}\bigg \lvert_{x_l=x_{l-}}^{x_{l+}} \ .
\end{equation}
The sums over $\mu$ become $\delta$-functions at the boundary using the Poisson summation formula, exactly as in the one-dimensional case \cite{BachelorStudents},
\begin{equation}
    \sum_k \int_a^b dx\, f(x)e^{i k (x-a)} = (b-a) f(a) \ .
\end{equation}
Applying this to our case we see that
\begin{equation}
    \sum_\mu \expval{\Psi | \Phi_{\vec{y}_0, \mu}} \Phi_{\vec{y}_0, \mu,+}^*(x_l)  \bigg \lvert_{x_{l-}}^{x_{l+}} =  \frac{1}{2} \Psi^*(x_l, \vec{y}_0)\bigg \lvert_{x_l=x_{l-}}^{x_{l+}} \ .
\end{equation}
Therefore
\begin{multline}
    \expval{(\hat{m} \cdot \vec x)(-i \hat{l} \cdot \vec \nabla)} = \expval{(\hat{l} \cdot \vec p_R) (\hat{m} \cdot \vec{x})}+i (\hat{l} \cdot \hat{m}) \\
    -\frac{i}{2}\int d^{d-1}\vec{y}_0 \, (\hat{m} \cdot \vec x) \Psi^*(x_l, \vec{y}_0) \Psi(x_l, \vec{y}_0)\bigg  \lvert_{x_l=x_{l-}}^{x_{l+}} \ .
\end{multline}
This last integral is again an integral over the surface $\p\Omega$ where we consider only the component parallel to $\hat{l}$. This can therefore be written in its final form as,
\begin{equation}
    \expval{(\hat{m} \cdot \vec x)(-i \hat{l} \cdot \vec \nabla)} = \expval{(\hat{l} \cdot \vec p_R) (\hat{m} \cdot \vec{x})}+i (\hat{l} \cdot \hat{m})-\frac{i}{2}\int_{\p\Omega} d^{d-1}\vec{x} \, ( \vec n \cdot \hat{l}) (\hat{m} \cdot \vec x) \abs{\Psi}^2 \ ,
\end{equation}
which is equivalent to
\begin{equation}
    \expval{(\hat{m} \cdot \vec x)(-i \hat{l} \cdot \vec \nabla)} = \expval{(\hat{l} \cdot \vec p_R) (\hat{m} \cdot \vec{x})}+i (\hat{l} \cdot \hat{m})-\frac{i}{2}\expval{( \vec n \cdot \hat{l}) (\hat{m} \cdot \vec x)}_{\p\Omega} \ .
\end{equation}
These correctly reproduce their one-dimensional versions. We're finally ready to plug everything into the generalized uncertainty relation eq.\eqref{eq:general inequality}. Calling $(\Delta B_m)^2 \equiv (\Delta x_m)^2$, we find
\begin{multline}
    2m \expval{T_k} \geq \frac{1}{(\Delta x_m)^2 } \bqty{\tfrac12 \expval{ \{(\hat{k} \cdot \vec p_R), x_m\}} - \expval{x_m} \expval{\hat{k}\cdot \vec p_R} }^2 +\\ +\frac{1}{4 (\Delta x_m)^2 } \bqty{(\hat{k} \cdot \hat{m}) - \expval{ (\vec n \cdot \hat{k}) (\hat{m} \cdot \vec x)}_{\p\Omega} + \expval{x_m}\expval{\vec{n} \cdot \hat{k} }_{\p\Omega} }^2+\\
    + \expval{ (\vec n \cdot \hat{k}) (\hat{k}\cdot \vec \nabla) }_{\p\Omega} + \expval{\hat{k} \cdot \vec p_R}^2 + \expval{\hat{k} \cdot \vec p_I}^2 \ ,
\end{multline}
which is again an inequality for the kinetic energy $T_k$. However, it contains a term, $\expval{ (\vec n \cdot \hat{k}) (\hat{k}\cdot \vec \nabla) }_{\p\Omega}$, which is not necessarily measurable. Hence we sum over a set of orthogonal directions $k$, which leads to
\begin{multline}
    \label{eq:measurable inequality}
    2m \expval{T} \geq \frac{1}{(\Delta x_m)^2 } \sum_{\hat{k}} \bqty{\tfrac12 \expval{ \{(\hat{k} \cdot \vec p_R), x_m\}} - \expval{x_m} \expval{\hat{k}\cdot \vec p_R} }^2 +\\ +\frac{1}{4 (\Delta x_m)^2 } \sum_{\hat{k}} \bqty{(\hat{k} \cdot \hat{m}) - \expval{ (\vec n \cdot \hat{k}) (\hat{m} \cdot \vec x)}_{\p\Omega} + \expval{x_m}\expval{\vec{n} \cdot \hat{k} }_{\p\Omega} }^2+\\
    + \expval{\gamma}_{\p\Omega} + \expval{\vec p_R}^2 + \expval{\vec p_I}^2 \ ,
\end{multline}
for any choice of direction $\hat{m}$, where we used the boundary conditions eq.\eqref{eq:robinbc}. Each term in eq.\eqref{eq:measurable inequality} is, in principle, measurable and therefore the inequality provides a physically meaningful interpretation of the Heisenberg uncertainty principle for the operator $-i \vec \nabla$ as an inequality for the kinetic energy of the system. This is only possible because of the introduction of the new momentum concept $\vec{p} = \vec{p_R}+i \vec{p_I}$, which is a measurable observable, and therefore reinforces the notion that the new momentum concept is the appropriate notion of momentum for a particle confined in a finite region of space. 

\section{Conclusions}

The usual momentum operator $-i\vec \nabla$ for a particle confined in a finite region of space is not self-adjoint, and therefore does not qualify as a physically valid observable. Based on the construction of a self-adjoint momentum operator for a particle in a finite one-dimensional interval, first introduced in \cite{alHashimiWieseAltMomentum, alHashimiWieseHalfLine}, we extended the new momentum concept to a finite region in arbitrary dimension. The new momentum concept provides an observable momentum which may be used to perform momentum measurements and compute expectation values, which would not be possible with the usual formulation. We have extended several results first obtained in the one-dimensional case \cite{BachelorStudents}, such as the Ehrenfest theorem and the interpretation of the Heisenberg uncertainty relation. 

The introduction of the new momentum $\vec{p} = \vec{p_R} + i\vec{p_I}$ provides an extension of the fundamental physical concept of momentum to the case of a quantum mechanical particle confined to a finite region of space. A central result of the present work is that, in a finite region, momentum should only be considered one direction at a time. The most striking manifestation of this fact is that different components of the momentum cannot in general be measured simultaneously.

Several remaining questions deserve further attention. Among these is understanding the dependence on the ultraviolet details of the probabilities of measurement of momentum eigenvalues, in both one and higher dimensions. In fact, a momentum measurement transfers infinite energy to the particle and therefore leads it outside the physical space. As such, what happens to the particle after a measurement necessarily depends on the underlying ultraviolet details. On the other hand, one would hope that measurement probabilities only depend on the low-energy physics.

Moreover, while the new momentum concept is in principle measurable, it would be interesting to construct a momentum measurement device, at least a theoretical one, along the lines first established by von Neumann \cite{VonNeumann32}. This could be done, for example, via time-of-flight measurements \cite{TimeOfFlight}. It would then be especially interesting to look for experimental verification of both the quantum mechanical force at the boundary (which arises as part of the momentum-force Ehrenfest theorem \eqref{eq:momentum force ehrenfest theorem}) and the interpretation of the Heisenberg uncertainty relation \eqref{eq:measurable inequality}.

\section*{Acknowledgments}

UJW thanks M.\!\! Al-Hashimi for his collaboration on the development of the new momentum concept in \cite{alHashimiWieseAltMomentum, alHashimiWieseHalfLine}. The research leading to these results received funding from the Schweizerischer Nationalfonds (grant agreement number 200020\_200424).

\bibliographystyle{ieeetr}
\bibliography{biblio}

\end{document}